\documentclass[12pt]{article}

\usepackage[english]{babel}
\usepackage[usenames]{color}
\usepackage[cp1250]{inputenc}
\usepackage{amsfonts}
\usepackage{amsthm}
\usepackage{graphicx}
\usepackage{epsfig}

\theoremstyle{plain}

\begin{document}
\newcommand{\bea}{\begin{eqnarray}}
\newcommand{\eea}{\end{eqnarray}}
\newcommand{\be}{\begin{equation}}
\newcommand{\ee}{\end{equation}}
\newcommand{\beas}{\begin{eqnarray*}}
\newcommand{\eeas}{\end{eqnarray*}}
\newcommand{\bs}{\backslash}
\newcommand{\bc}{\begin{center}}
\newcommand{\ec}{\end{center}}

\title{Optimal compression of hash-origin prefix trees}
\author{Jarek Duda}

\date{\it \footnotesize Jagiellonian University, Cracow, Poland, \\
\textit{email:} dudaj@interia.pl}
\maketitle

\begin{abstract}
There is a common problem of operating on hash values of elements of some database. In this paper there will be analyzed informational content of such general task and how to practically approach such found lower boundaries. Minimal prefix tree which distinguish elements turns out to require asymptotically only about 2.77544 bits per element, while standard approaches use a few times more. While being certain of working inside the database, the cost of distinguishability can be reduced further to about 2.33275 bits per elements. Increasing minimal depth of nodes to reduce probability of false positives leads to simple relation with average depth of such random tree, which is asymptotically larger by about 1.33275 bits than $\lg(n)$ of the perfect binary tree. This asymptotic case can be also seen as a way to optimally encode $n$ large unordered numbers - saving $\lg(n!)$ bits of information about their ordering, which can be the major part of contained information. This ability itself allows to reduce memory requirements even to $\ln(2)\approx 0.693$ of required in Bloom filter for the same false positive probability.
\end{abstract}

\section{Introduction}

There is often considered a problem of counting the size of some discrete family, like the number of full binary trees (all nodes have 0 or 2 children) with $n$ leaves is the Catalan number (\cite{conc}): $C_n={2n \choose n}/(n+1)$, which is asymptotically $\frac{2^{2n}}{n^{3/2}\sqrt{\pi}}$. Choosing one of $m$ elements requires $\lg(m)$ bits of information ($\lg\equiv \log_2$), so the minimal average amount of information to choose one of such trees is asymptotically $2$ bits per leaf.

The situation usually improves if the probability distribution of elements is nonuniform: if it is $\{p_i\}_{i=1..n}$, the average number of bits per element required to encode such proportions of elements is the Shannon entropy:
\be \left(\lim_{N\to\infty} \frac{1}{N}\lg\left(\frac{N!}{(p_1N)!(p_2N)!..(p_nN)!}\right)=\right)\qquad-\sum_{i=1}^n p_i \lg(p_i)\leq \lg(n)\qquad\ee
where the equality holds only for uniform distribution. This formula can be seen that we need $\lg(1/p_i)$ bits of information to encode choice/symbol of $p_i$ probability and so at average we need Shannon entropy bits of information per choice. This informational capacity can be easily approached as near as we need using entropy coder, like Arithmetic Coding \cite{ari} operating on two states (boundaries of range), or recent single state Asymmetric Numeral Systems \cite{ans}, in which encoding $p_i$ probability symbol increases the state about $1/p_i$ times.
\begin{figure}[t!]
    \centering
        \includegraphics{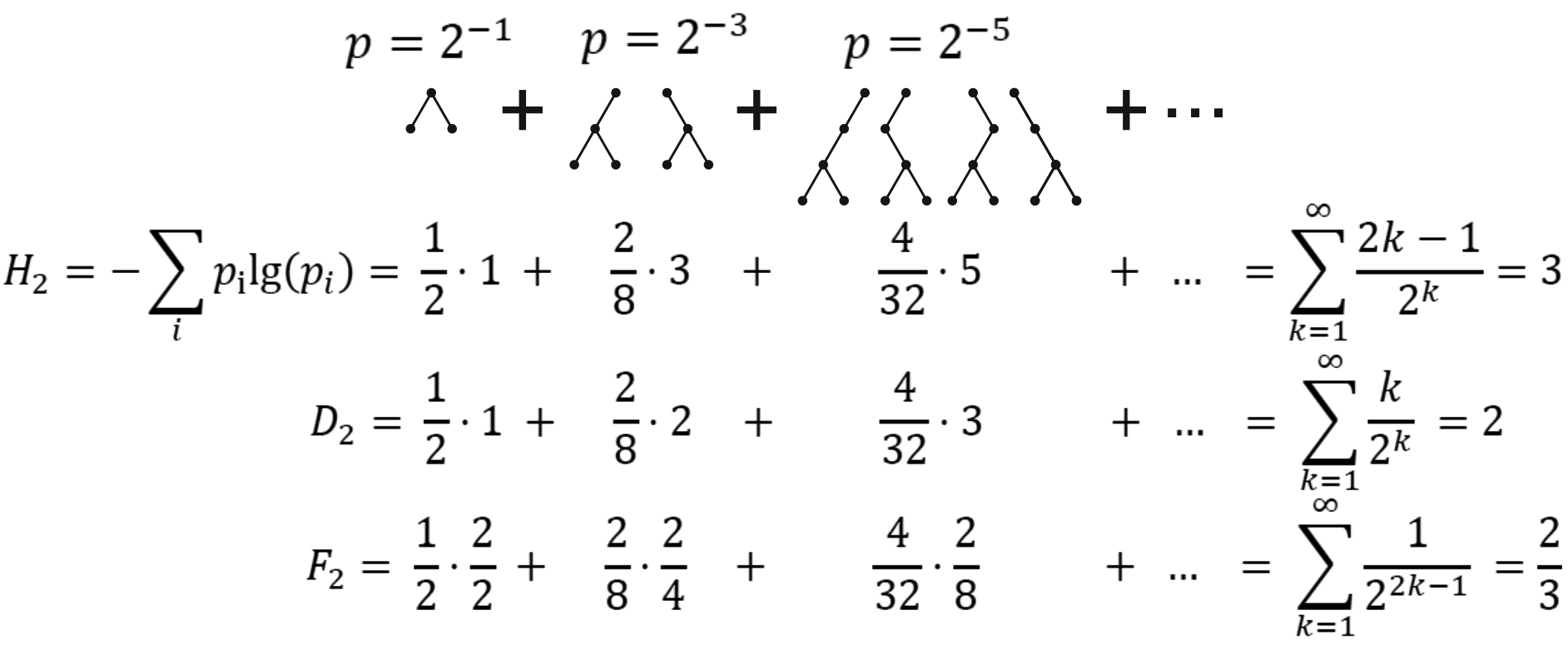}
        \caption{Calculation of average entropy ($H_2$), average depth ($D_2$) and false positive ($F_2$) probability (of accidently getting to a leaf) of infinite prefix tree family with two leaves.}
       \label{h1}
\end{figure}

The improvement of using probabilities while encoding becomes crucial if the set of possibilities is infinite. Using entropy coding would still lead to that for any length, there are elements requiring even more bits to store them. However, their probabilities may quickly drop, such that the average amount of required bits remains finite. In Fig. \ref{h1} we can see that we have such situation while considering the space of minimal prefix trees required to distinguish between some bit sequences, chosen with uniform distribution for each bit. \\

Such prefix tree is a natural representation while considering hash function of elements of a dictionary/database - deterministically chosen pseudorandom bit sequence assigned to each element. These sequences have usually fixed length, but we will not use such assumption here - we can imagine that initially they are infinite, then they are cut to individually chosen optimal finite length prefix: minimal to distinguish from others.

\begin{figure}[t!]
    \centering
        \includegraphics{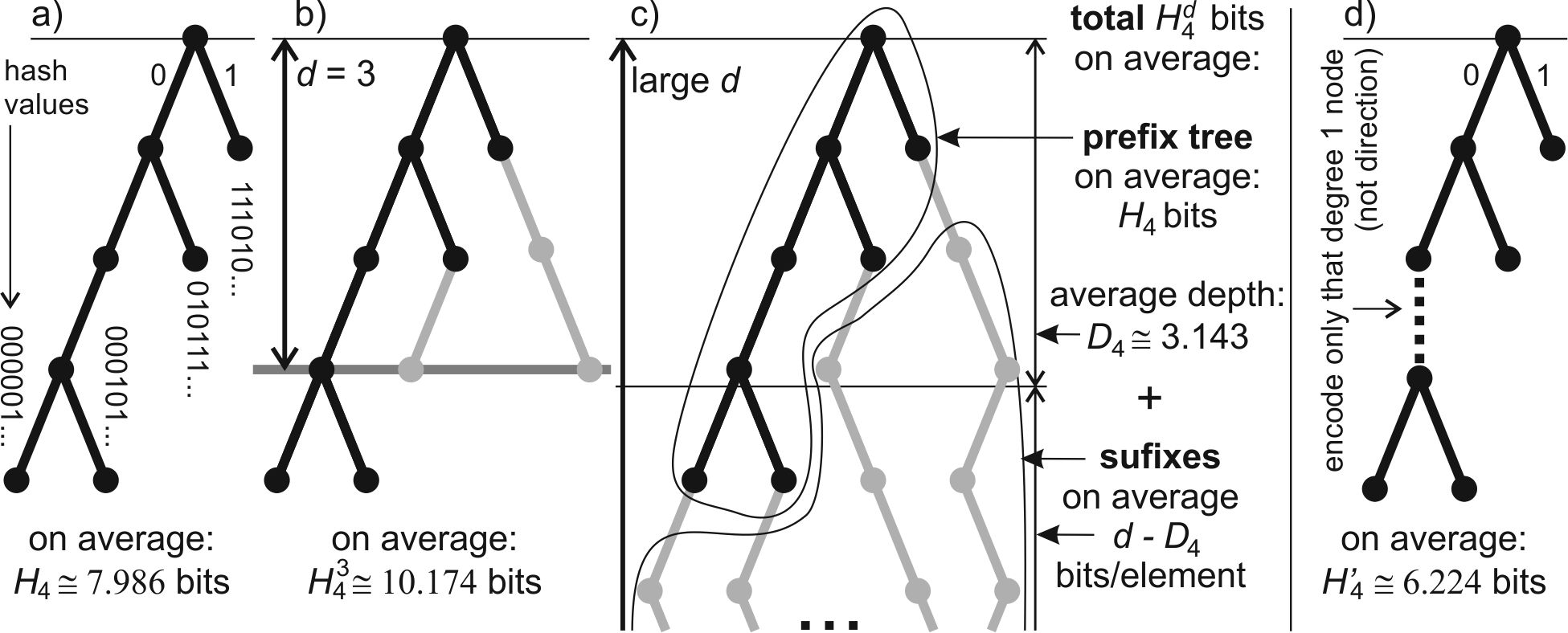}
        \caption{Three basic cases we will consider: a) minimal prefix tree, b) minimal depth $d$ prefix tree, c) asymptotic behavior for large $d$ and d) reduced minimal prefix tree.}
       \label{h2}
\end{figure}

So let us assume that there are $n$ infinite sequences of independently chosen bits with $P(0)=P(1)=1/2$ probability and we build the minimal prefix tree distinguishing these sequences, like in Fig. \ref{h2}a. Imagine that someone have only such a tree. While asking for some element from the dictionary, he would always get answer that it is in the tree, finding the corresponding leaf - there is not possible \emph{false negative} case. However, while asking for an element outside the dictionary, there is a nonzero probability to also get to a leaf - there are possible \emph{false positive} cases. By elongating paths to the leaves to some fixed minimal depth like in Fig. \ref{h2}b, we can use additional information about the sequences to reduce this probability as much as we need. Finally increasing this depth to some fixed large number, we would almost surely just store fixed length bit sequences, but without information about their order - it allows to save $\lg(n!)$ bits of information about the order.

We will see that statistical ensemble of minimal prefix trees generated this way has asymptotically Shannon entropy 2.77544 bits per element - this is the minimal average amount of information required to store such tree. These minimal trees have large false positive probability (about 0.721), but they can be safely used while there is certainty of working in the corresponding dictionary. Having this certainty, we can reduce the tree further to about 2.33275 bits/element, increasing false positive probability to 1. Attaching some additional information to the leaves of such tree allows to define their type, for example to classify if given word is noun or verb while being certain of working within given dictionary. In such case, the minimal amount of required bits per element is 2.33275 plus attached information, like 1 bit to tell if the word is verb or noun (or less if they have unequal probabilities). The minimality of used information can be also seen as advantage for cryptographic purposes, reducing costs of eventual leaks of such information.

Considered data compression is rather impractical while operating on the tree, but it might be useful while storing or transmitting it. Additionally, these considerations and known theoretical boundaries may help developing methods to handle memory in a more optimal way.

Beside informational content, there will be also found average depth of such random tree and we will relate values of these properties. These relations lead to relatively compact formulas for some complicated recurrences. We will also find probability of false positives and use it to compare this approach with commonly used Bloom filter. Table \ref{tb} gathers calculated properties for some more interesting parameters.
\section{Entropy of minimal prefix tree}
The additivity of entropy allows us to divide the selection of element into a smaller choices. To encode such randomly generated prefix tree, we can divide the choice of the tree into situations in single nodes. It is essential to properly calculate the probabilities of choices, in such case all representations has to lead to the same entropy. There are plenty of possibilities, like just storing if the node has none, left, right or both children.

For convenience we will use different representation here: for each node store distribution of sequences between its left and right child. Specifically, assume that for given node we know the total number ($n$) of sequences going through it (leaves in its subtree). Now for this node we will only encode how many of these sequences make the next step left ($k\in \{0,1,..,n\}$) - the rest of them ($n-k$) go right. We can then recursively repeat this process for both children, knowing the total number of leaves in their subtrees. Imagining bit sequences as infinite binary expansions of numbers from $[0,1]$ range, we can see it as that the first choice is how many of these numbers goes to the $[0,1/2]$ half, then we recursively go into the two subranges.

There is a problem with the root of the tree - the total number of elements has to be written separately. This issue will not be addressed here, but storing a natural number takes about $\lg(n)$ bits of information, so this cost divided by the number of elements vanishes asymptotically.\\

Denote by $H_n$ the average number of bits required to encode minimal prefix tree for $n$ bit sequences (leaves), which bits were independently randomly chosen with $P(0)=P(1)=1/2$ probability distribution. Connecting the node with its two children we get the recurrence:
$$ H_n = \sum_{k=0}^n \frac{{n \choose k} }{2^n} \left(-\lg\left(\frac{{n \choose k}}{2^n}\right)+H_k+H_{n-k}\right)=h_n + 2\sum_{k=0}^{n-1} \frac{{n \choose k} }{2^n} H_k  +2\frac{H_n}{2^n}$$
where $h_n := -\sum_{k=0}^n \frac{{n \choose k} }{2^n} \lg\left({n \choose k}/2^n\right)$.\\
Subtracting $2\frac{H_n}{2^n}$ from both sides and then dividing by $(1-1/2^{n-1})$ we finally get:
\be H_n= \frac{ 2^{n-1}h_n+\sum_{k=0}^{n-1} {n \choose k} H_k }{2^{n-1}-1} \qquad \mathrm{for}\ n\geq 2, \qquad \qquad H_0=H_1=0 \label{hn}\ee
For node through which $n$ sequences go, the $h_n$ is the average amount of information required to choose how many of them make the next step left.  Practical encoding of the tree using this approach requires to go through all internal nodes in some order (e.g. preorder) and encode this information for each of them (and eventually some information stored in leaves).

In practice, such single choice of one of $n+1$ possibilities can be divided into a few smaller choices, like if $k<n/2$ as the first one. Using binary choices and choosing the divisions to make probabilities near $1/2$ would allow to straightforward encode these choices as bits of information in such created e.g. Huffman tree. However, these probabilities are not exactly powers of 2, making that we would get away from the theoretical informational capacity this way. Using precise entropy coder instead, like Arithmetic Coding or Asymmetric Numeral Systems, allows to easily get as near the calculated boundary as we want.\\

The recurrence (\ref{hn}) for $H_n$ seems to be very difficult to solve. We will find more analytical formula later thanks of relating it with average depth. Let us now find some its approximation. The probabilities in the $h_n$ formula are nearly as for the Gaussian distribution - we can approximate it as:
\be \tilde{h}_n := -\int_{-\infty}^{\infty} \rho_{n/2,\sqrt{n/4}}(x) \lg\left(\rho_{n/2,\sqrt{n/4}}(x)\right)\, dx = \frac{1}{2}\lg\left(\frac{\pi e}{2}n\right) \ee
where $\rho_{\mu,\sigma}(x):=\frac{1}{\sqrt{2\pi}\sigma} e^{-\frac{(x-\mu)^2}{2\sigma^2}}$ is the Gaussian distribution probability density.

The next approximation is using that Gaussian distribution for large $n$ is almost completely concentrated in the center, leading to much simpler recurrence which can be easily solved:
$$ \tilde{H}_n=\tilde{h}_n+2\tilde{H}_{n/2} = \frac{1}{2}\lg\left(\frac{e\pi}{2}n\right) +2\tilde{H}_{n/2} $$
\be \tilde{H}_n = \alpha n - 1 - \frac{1}{2}\lg\left(\frac{e\pi}{2}\right) - \frac{1}{2}\lg(n) \ee
The recurrence for $\tilde{H}_n$ leaves a freedom of choosing the most important parameter here: $\alpha$, which is the average number of bits per element in this nearly linear relation. We can fit it to the numerical solutions of (\ref{hn}) for now, but later we will see suggestion to choose this average cost of distinguishability as $\alpha=\frac{1}{2}+(1+\gamma)\lg(e) \approx 2.77544121816583$.

\begin{figure}[t!]
    \centering
        \includegraphics{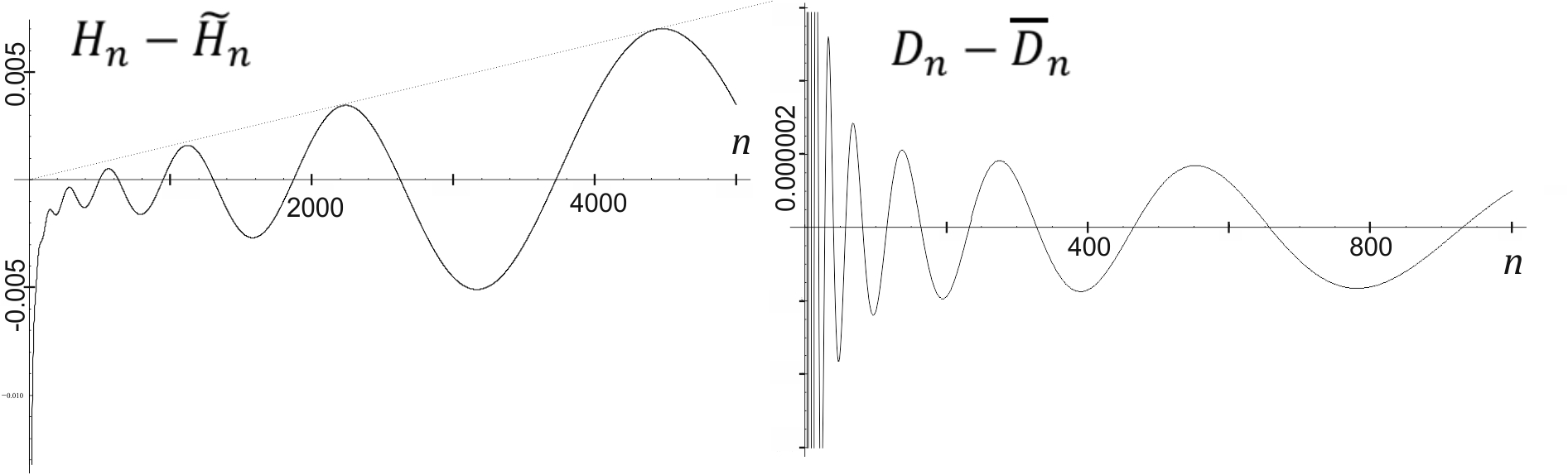}
        \caption{Oscillations of $H_n$ and $D_n$ using different approximations.}
       \label{h3}
\end{figure}
$\tilde{H}_n$ seems to approximate $H_n$ well, but surprisingly there appear some small regular oscillations in their difference, what can be seen in Fig. \ref{h3}. Looking at $f(x)=2f(x/2)$ functional equation (nearly as for $\tilde{H}_n$), we can understand the origin of this difference - there can appear additional growing oscillating $f(x)=x\sin(2\pi \lg(x))$ solution in positive half axis. Finally
\be \overline{H}_n := \alpha n - 1 - \frac{1}{2}\lg(e\pi/2) - \frac{1}{2}\lg(n) + \epsilon n\sin(2\pi \lg(n)+\delta)\ee
approximates $H_n$ for larger $n$ with precision about $10^{-5}$ for fitted $\epsilon$, $\delta$:
\be \alpha= 2.7754412\qquad\qquad \epsilon=1.6\cdot 10^{-6}\qquad\qquad \delta=0.88 \label{albe}\ee
The last growing oscillating term makes that the average bits of information per element is asymptotically in width $2\epsilon$ range $[2.7754396,2.7754428]$.

\section{Minimal depth prefix tree}
The minimal prefix tree allows to distinguish between elements, for example to attach some information to them. However, if we are interested in question if given object is in the dictionary, such tree would often falsely give a positive answer for any random element (about 0.721 probability as we will see later). We can reduce this false positive probability by elongating leaves further as in Fig. \ref{h2}b,c - at cost of storing these suffixes up to some chosen depth $d$. We assume that these sequences are completely random, so it costs 1 bit per node.

While calculating average informational content of such minimal depth $d$ prefix tree, the only change from the minimal prefix tree is the cost of leaves - now it grows from 0 to the remaining depth. Expressing it in the language of recurrence:
\be H^0=H,\ H^d_1=d,\qquad \textrm{for}\ d\geq 1,\ n\geq 2:\quad H^d_n = h_n + \sum_{k=1}^{n} \frac{{n \choose k} }{2^{n-1}} H^{d-1}_k \ee
for growing $n$, leaves will statistically get over this boundary: $\lim_{n\to \infty} H^d_n-H_n=0$.

This recurrence differs from the one for $H_n$ only by $H^d_1=d$ instead of $0$. We can trace the increase of the original $H_n$ which came from such additional 1 from depth $d$ to write:
\be H^d_n=H_n+L^{d-1}_n+2L^{d-2}_n+..+(d-1)L^1_n=H_n+\sum_{i=1}^{d-1} i L^{d-i}_n \label{Hd}\ee
where this $L$ fulfills recurrence:
$$ L^0_1=1,\ \forall_{i>1} L^0_i=0, \qquad \mathrm{for}\ d\geq 1,\ n\geq 1:\quad L^d_n =\sum_{k=1}^{n} \frac{{n \choose k} }{2^{n-1}} L^{d-1}_k $$
Equation (\ref{Hd}) brings natural interpretation of $L^d_n$, which is confirmed by the recurrence: it is just the expected number of depth $d$ leaves of the minimal prefix tree. It means $l^d_n:=\frac{L^d_n}{n}$ is the probability that a hash value will correspond to a depth $d$ leaf, so $\sum_{d=0}^{\infty} l^d_n =1$. We will find analytic formula for $l^d_n$ later.
\section{Asymptotic behavior and average depth}
Let us denote by $D_n$ the average depth of leaves in size $n$ minimal prefix tree:
\be D_n = \sum_{d=0}^\infty d\, l_n^d \ee
Situation for large $d$ looks like in Fig. \ref{h2}c - practically none of leaves get below the boundary. We can see encoding of such tree as storing the minimal prefix tree plus at average $d-D_n$ bits per element to encode suffixes, like in this figure. So for large $d$, $H_n^d$ goes to $H_n + n(d-D_n)$ and they are equal in the limit. It can be also seen from (\ref{Hd}) formula. Alternative view of encoding this information (without leaves below $d$) is that it is just encoding $n$ numbers in $[0,2^d-1]$ range - one of ${2^d \choose n}$ combinations, which for large $d$ is practically $\frac{2^{dn}}{n!}=2^{dn-\lg(n!)}$:
 \be \textrm{for large }d >\lg(n)\ :\qquad\qquad H^d_n\approx dn-\lg(n!)\ee
The amount of information does not depend on the choice of one of equivalent representations. Finally, from that $H_n + n(d-D_n)$ is asymptotically equal to $dn-\lg(n!)$, we can remove the dependence from $d$, getting simple relation:
\be H_n+\lg(n!)=n D_n \label{rel}\ee
This equality can be also seen straightforward: on its both sides there is the amount of bits required to store the minimal prefix tree including information about the order of leaves.

Let us now find $D_n$ in alternative ways now, like through recurrence. $D_0=D_1=0$,
$$D_n=\sum_{k=0}^n \frac{{n \choose k}}{2^n} \frac{k([k>0]+D_k)+(n-k)([n-k>0]+D_{n-k})}{n}=
\sum_{k=1}^{n-1} \frac{{n \choose k}}{2^{n-1}} \frac{k}{n} (1+D_k)+\frac{1+D_n}{2^{n-1}}$$
\be D_n=\frac{\sum_{k=1}^{n-1} {n-1 \choose k-1} (1+D_k)+1}{2^{n-1}-1}=\frac{2^{n-1}+\sum_{k=2}^{n-1} {n-1 \choose k-1} D_k}{2^{n-1}-1}\qquad \textrm{for}\ n\geq 2\ee
where $[c]=1$ if condition $c$ is true and 0 otherwise. Using the (\ref{rel}) relation and Stirling approximation $n!\approx \sqrt{2\pi n} \left(\frac{n}{e}\right)^n$, we get
\be D_n\approx \tilde{D}_n:= \lg(n)+ 1.332746156 - \frac{\lg(e)}{2n} + \epsilon \sin(2\pi \lg(n)+\delta) \ee
In the case of creating perfect binary tree: that $n=2^m$ prefixes would take all $[0,2^m-1]$ values, the average depth would be exactly $m=\lg(n)$. We see that the randomness of the tree increases this average depth asymptotically by $1.332746156=\alpha-\lg(e)$.\\

Let us now find $l_n^d$: the probability that a sequence will correspond to depth $d$ leaf. Taking a sequence, it corresponds to depth $d$ leaf if among the other $n-1$ sequences, there is $k\geq 1$ having the same $d-1$ length prefix, but none of them have the same $d$ length prefix:
\be l^d_n=\sum_{k=1}^{n-1} {n-1 \choose k} (2^{-d+1}-2^{-d})^k (1-2^{-d+1})^{n-1-k}=(1-2^{-d})^{n-1}-(1-2^{-d+1})^{n-1} \label{ddn}\ee
We can for example use it to calculate moments, like the average depth we required:
$$D_{n+1}=\sum_{d=1}^{\infty} d\left((1-2^{-d})^n-(1-2^{-d+1})^n\right)=\sum_{d=1}^{\infty} d\sum_{k=1}^n {n \choose k}\left((-2^{-d})^k - (-2^{-d+1})^k\right)=$$
$$=\sum_{k=1}^n {n \choose k} (-1)^k (1-2^k)\sum_{d=1}^{\infty} 2^{-dk} d=\sum_{k=1}^n {n \choose k} (-1)^k (1-2^k) \frac{2^k}{(2^k-1)^2}=
-\sum_{k=1}^n {n \choose k} \frac{(-1)^k}{1-2^{-k}}$$
The sign change makes this form still inconvenient - let us expand it further:
$$D_{n+1}=-\sum_{k=1}^n {n \choose k} (-1)^k \sum_{i=0}^{\infty} \left(2^{-i}\right)^k=-\sum_{d=0}^{\infty} \sum_{k=1}^n {n \choose k} \left(-2^{-d}\right)^k $$
\be D_{n+1} =\sum_{d=0}^{\infty} 1 - (1-2^{-d})^n \label{dnf}\ee
This formula can be also understood straightforward. $(1-2^{-d})^n$ is the probability that no sequence will reach some fixed length $d$ sequence - we can for example use it to find saturation of possible $2^d$ internal nodes on depth $d$ (reached by at least two sequences): the expected number of depth $d$ internal nodes is
 $$2^d \left(1-(1-2^{-d})^n-n 2^{-d}(1-(1-2^{-d})^{n-1}\right)$$
Fixing a sequence, $1 - (1-2^{-d})^n$ term from (\ref{dnf}) can be seen as the probability that some other sequence will have the same length $d$ prefix. So to distinguish our sequence, it has to contain edge corresponding to the succeeding step. Finally the (\ref{dnf}) formula can be seen as the sum of probabilities of that the path to the leaf contains $[d,d+1]$ edge.\\

The $f_n(x):=1 - (1-2^{-x})^n$ function is kind of step-like decreasing function for $x\in[0,\infty)$: nearly $1$ up to about $\lg(n)$, then quickly drops to nearly $0$ and stays there. It allows to understand the source of the oscillations: the exact sum depends on relative position of this drop to the discrete lattice of natural numbers. We could try to use Euler-Maclaurin formula (\cite{conc}): change summation into integration. 
$$\int_{0}^{\infty} 1-(1-2^{-x})^n dx = \lg(e) \int_{0}^{1} \frac{1-u^n}{1-u} du = \lg(e) \int_{0}^{1} 1+u+..+u^{n-1} du =\lg(e)\sum_{k=1}^n\frac{1}{k}$$
where we have used $u=1-2^{-x}$ substitution. The initial Euler-Maclaurin approximation is

\be\overline{D}_{n+1}:=\frac{f_n(0)}{2}+\int_0^\infty f_n(x) dx =\frac{1}{2}+\lg(n)+\lg(e)\left(\gamma+\frac{1}{2n}-\frac{1}{12n^2}+\frac{1}{120n^4}-...\right) \ee
The Euler-Maclaurin formula requires also sum of series of derivatives in both boundaries. All derivatives vanish in infinity, however in $0$ only up to $(n-1)$-st are zero ($f_n^{(m)}(0)=-(-1)^{n+m}n! (\ln(2))^m  S(m,n)$ where $S(m,n)$ are Stirling numbers of the second kind). Unfortunately from Fig. \ref{h3} we see that oscillations are exploding in $0$ (and then decreases to constant level), making the Euler-Maclaurin series divergent in 0. However, the integral suggests the choice of the basic parameter for $H_n$ and $D_n$:
\be \alpha-\lg(e)=\lim_{n\to \infty} \overline{D}_n - \lg(n) = \frac{1}{2}+\gamma \lg(e) \approx 1.3327461772768672 \ee

\section{False positive probability and comparison with Bloom filter}
We can now calculate probability of false positive cases - that a completely random sequence will get to a leaf. Let us denote this probability by $F_n$ for the minimal prefix tree and $F_n^d$ for the minimal depth $d$ tree. Recurrence relations for them are
$$F_0=0,\ \ F_1=1,\ \ F_n=\sum_{k=0}^n  \frac {{n \choose k}}{2^n} \frac{F_k+F_{n-k}}{2}= \sum_{k=1}^{n-1}  \frac {{n \choose k}}{2^n} F_k+\frac{F_n}{2^n}$$
$$ F^0_n=F_n=\frac{\sum_{k=1}^{n-1} {n \choose k} F_k}{2^n-1}\qquad\qquad F^{d+1}_n= \sum_{k=1}^n  \frac {{n \choose k}}{2^n} F^d_k $$
Having probability of leaf's depth (\ref{ddn}), we can also find straightforward formulas:
$$F_n=n\sum_{d=1}^{\infty} l^d_n\, 2^{-d}=n\sum_{d=1}^{\infty} 2^{-d}\left((1-2^{-d})^{n-1}-(1-2^{-d+1})^{n-1}\right)=\frac{n}{2}\sum_{d=1}^{\infty} 2^{-d}(1-2^{-d})^{n-1}$$
$$F^l_n = 2^{-l} n \sum_{d=1}^{l} l^d_n + n\sum_{d=l+1}^{\infty} l^d_n\, 2^{-d}=\frac{n}{2}\sum_{d=l}^{\infty} 2^{-d}(1-2^{-d})^{n-1}$$
We can approximate this sum by integral (calculated using $u=1-2^{-x}$ substitution):
$$F_n\approx \frac{n}{2} \int_{x=0}^{\infty} 2^{-x}(1-2^{-x})^{n-1} dx=\frac{n\lg(e)}{2} \int_{u=0}^1 u^{n-1} du=\frac{\lg(e)}{2}~\approx 0.72134752$$
There appear $\sin(2\pi \lg(n))$ type oscillations like previously - while in a perfect tree all sequences would get to a leaf, for the minimal prefix trees $F_n$ turns out to oscillate  between 0.72134 and 0.72136. As expected, for $n \ll 2^d$, $F_n^d$ grows approximately like $\frac{n}{2^d}$, then saturates near $n=2^d$ and finally oscillates in the above range.\\

Let us compare required amount of information with commonly used Bloom filter \cite{bloom}. In this method we use length $m$ bit table initialized with zeroes. For each element there are calculated $k$ independent hash values from $[1,m]$ range - positions in the table. Inserting the element is changing values in all these $k$ positions to "1". The question if there is stored given element is asking if corresponding $k$ positions are "1".

As in the presented approach, false negative cases are not possible. False positive case appears when accidently all $k$ positions are set to "1" by some of $n$ elements - probability of this situation is
$$p_f=\left(1-\left(1-\frac{1}{m}\right)^{kn}\right)^k\approx \left(1-e^{-kn/m}\right)^k$$
for given $n$, $m$, this probability is minimized for $k=\frac{m}{n}\ln(2)$. As we should expect, this $k$ corresponds to the case that bits in all positions of the table has exactly $p=1/2$ probability. We assume that they are uncorrelated - in this case the table contains maximal amount of information ($m\left(-p\lg(p)-(1-p)\lg(1-p)\right)=m$ bits of information) and so cannot be further compressed.

This optimal choice of $k$ leads to the false positive probability $p_f=e^{-\frac{m}{n} (\ln(2))^2}$. So for chosen $p_f$, the optimally chosen $k$ means the Bloom filter memory requirements is
\be (m=)\ B_n(p_f):= -\frac{n\lg(p_f)}{\ln(2)} \qquad\qquad \textrm{bits of information}\ee
which cannot be compressed further.

To compare with presented approach, small $p_f$ is approximately $n/2^d$, so we should choose $d=\lg(n/p_f)$. For large $d$, $H^d_n\approx dn-\lg(n!)$, so in analogous situation we need approximately
\be H^{\lg(n/p_f)}_n \approx n \lg(n/p_f) -n! \approx - n\lg(p_f)+n\lg(e)\ee
so asymptotically the ability to save $\lg(n!)$ bits of information about the order of hash values itself, allows to reduce memory requirements even to  $\ln(2)\approx 0.693$ of used in Bloom filter.\\

Table \ref{tb} contains comparison of memory requirements of both methods - Bloom filter is better only for small $d$: when prefix tree is not intended to provide small false positive probability. In this case, in opposite to Bloom filter, it additionally contains information to distinguish all elements, for example to attach some additional information to them.

The large false positive probability is no longer a disadvantage for cryptographic purposes. Just oppositely - sometimes we would like to send/store as little as possible in case of obtaining it by a third party - by requiring to use shared secret information to make use of the message. The presented approach can be used for example when we share the same database and we would like to transmit some additional properties of its elements - often false positives would make it useless while not having the original database. The next section shows how to improve it further to the optimum.
\section{Reduced minimal prefix tree}
The fact that the minimal prefix tree has still some ability to exclude elements from the set (false positive probability is less than 1), suggests that the informational content can be further reduced while being ceratin of working inside the set. The idea was pointed me out by James Dow Allen: for internal nodes of degree 1, all sequences from the set choose the same direction, so we do not need to encode this direction - saving 1 bit of information per such node. These degree 1 nodes gave the minimal prefix tree the ability to sometimes (about $1-0.721$ probability) recognize that an element is not from the the set - presented reduction decreases this probability to 0 (false positive probability grows to $1$).

The average informational content of such $n$ leaf reduced tree ($H'_n$) can be calculated as for the minimal prefix tree, but using a bit smaller cost of encoding single step ($h'_n<h_n$). In corresponding way this reduced tree can be encoded in practice. This time for node through which $n$ sequences go, two boundary possibilities of $2^{-n}$ probability: that all these sequences make the succeeding step left/right, are merged into one situation of $2^{-n+1}$ probability - that degree of this node is 1:
$$h'_n := -\sum_{k=1}^{n-1} \frac{{n \choose k} }{2^n} \lg\left({n \choose k}/2^n\right) + (n-1)2^{-n+1}=h(n) - 2^{-n+1}$$
In other words, we save 1 bit per each appeared degree 1 node, so at average we save $2^{-n+1}$ bits per each node through which $n$ sequences go. Now $H'_n$ can be calculated using (\ref{hn}) recurrence for $H_n$, but with $h_n$ replaced by $h'_n$. Knowing that due to the reduction we save $2^{-m+1}$ bits for each node through which $m\leq n$ sequences go, allows to find straightforward formula for $H_n-H'_n$. The expected number of such nodes is the sum of their expected number on depth $d$:
$$N_n^m = [n=m] + \sum_{d=1}^{\infty} 2^d {n \choose m} 2^{-dm} (1-2^{-d})^{n-m}$$
\be H_n-H'_n=\sum_{m=2}^n 2^{-m+1}N_n^m =2^{-n+1}+\sum_{d=1}^{\infty} 2^d \sum_{m=2}^n 2^{-m+1} {n \choose m} 2^{-dm} (1-2^{-d})^{n-m} \ee
It is also the expected number of degree 1 nodes. Using the fact that the number of degree 2 nodes is $n-1$, we can find simpler formula. So let us calculate the expected number of all nodes and then subtract the expected number of remaining nodes.

The expected number of nodes on depth up to $k$ is $\sum_{d=0}^{k} 2^d\left(1-(1-2^{-d})^n\right)$. This number includes expansions of leaves - there are at average $n(k-D_n)$ of them for large $k$. Subtracting them and taking $k\to\infty$ limit, the expected number of degree 1 nodes is:
\be nD_n - \sum_{d=0}^{\infty} \left(n-2^d\left(1-(1-2^{-d})^n\right)\right)-(n-1) \label{d1n}\ee
where the subtracted $n-1$ term is the number of degree $2$ nodes. 

Let us approximate above sum with integral, using $u=1-2^{-x}$ substitution as previously:
$$\int_{0}^{\infty} n-2^x\left(1-(1-2^{-x})^n\right)dx=\lg(e)\int_{0}^1 \frac{n}{1-u}-\frac{1-u^n}{(1-u)^2} du=$$
$$=\lg(e)\int_{0}^1 \sum_{k=0}^{n-1}\frac{1-u^k}{1-u}du=\lg(e)\int_{0}^1 \sum_{i=1}^{n-1}(n-i)u^{i-1} du =\lg(e)\sum_{i=1}^{n-1}\frac{n-i}{i}$$
where we have used $\frac{1-u^n}{1-u}=\sum_{i=0}^{n-1}u^i$ twice, obtaining good approximation:
$$\sum_{d=0}^{\infty} \left(n-2^d\left(1-(1-2^{-d})^n\right)\right) \approx \frac{n}{2}+\lg(e)\left(n\sum_{i=1}^{n-1}\frac{1}{i}-n+1\right)-\lg(e/2)$$
the difference for this approximation is about $2\cdot 10^{-6}\cdot n\sin(2\pi\lg(n)-0.6)$. 

Now substituting $\overline{D}_n = \frac{1}{2}+\lg(e)\sum_{i=1}^{n-1}\frac{1}{i}$ to (\ref{d1n}), we get
\be nD_n - \sum_{d=0}^{\infty} \left(n-2^d\left(1-(1-2^{-d})^n\right)\right) - (n-1) \approx n\lg(e/2) \ee
So there is approximately $n\lg(e/2)\approx 0.442695n$ degree 1 nodes in $n$ leaf minimal prefix tree - the asymptotic expected number of bits per element required to encode such reduced tree is approximately $\alpha-\lg(e/2)=3/2+\gamma\lg(e)\approx 2.332746177$.
\section{Conclusions}
Presented analysis shows expected values and theoretical boundaries for naturally appearing prefix trees. Practical approach can easily reach these limits for example for various database applications. In this moment it could be used to optimally compress these data for transmission or storage purposes, but knowing these boundaries alone should motivate to search for online processing methods with more optimal memory usage.

If there is required small false positive probability, this approach requires asymptotically about $0.693$ of memory used by Bloom filter. From the other side, storing only the minimal prefix tree may have different applications, like classification (e.g. verb/noun) while being certain of working within some fixed dictionary. The ability of distinguishing elements allows to attach some information to them, paying for this ability at least additional 2.33275 bits per element. False positive probability equal 1 of such minimal send/stored information can be seen as additional desired property for cryptographic applications.\\

Another possible application is to optimally store unordered sequence of $n$ numbers - saving $\lg(n!)$ bits of information about their order. If these numbers densely cover e.g. some length $m$ range, we can create length $m$ bit table and mark their positions - optimal compression of such uncorrelated numbers would require ${n \choose m} \approx 2^{-nh(m/n)}$ bits of information, where $h(p)=-p\lg(p)-(1-p)\lg(1-p)$. However, the problem appears when $m$ is very large - in this case it might be more convenient to compress the prefix tree of these numbers as considered here: first encode the distribution between left and right halves of the range, then recursively go into these subranges.

\begin{table*}[t]
\caption{Values of considered functions for some parameters. $H$ is informational content in bits, $D$ average depth, $F$ probability of false positives and $B$ required bits of information using Bloom filter for analogous parameters. Storing $H^d_n$ cases in a standard way ($n$ length $d$ sequences) would require $n*d$ bits, while encoding the tree allows to save about $\lg(n!)$ bits choosing their ordering.}
\label{tb}
\centering
\begin{tabular}{|c||c|c|c|c|c|c|c|c|c|c|}\hline
$n$&1&2&3&4&5&6&7&8&9&10\\\hline\hline
$h'_n$       & 0     & 1     & 1.561 & 1.906 & 2.136 & 2.302 & 2.431 & 2.536 & 2.626 & 2.704  \\
$h_n$        & 1     & 1.5   & 1.811 & 2.031 & 2.198 & 2.333 & 2.447 & 2.544 & 2.630 & 2.706  \\
$\tilde{h}_n$& 1.047 & 1.547 & 1.840 & 2.047 & 2.208 & 2.340 & 2.451 & 2.547 & 2.632 & 2.708  \\
$H'_n$       & 0     & 2     & 4.082 & 6.224 & 8.407 & 10.62 & 12.84 & 15.08 & 17.34 & 19.60  \\
$H_n$        & 0     & 3     & 5.415 & 7.986 & 10.62 & 13.27 & 15.94 & 18.63 & 21.32 & 24.02  \\
$\overline{H}_n$&0.728& 3.004& 5.487 & 8.055 & 10.67 & 13.31 & 15.98 & 18.66 & 21.35 & 24.05  \\
$H^5_n$      & 5     & 9.125 & 12.79 & 16.15 & 19.30 & 22.31 & 25.19 & 27.99 & 30.72 & 33.39  \\
$H^9_n$      & 9     & 17.00 & 24.44 & 31.46 & 38.17 & 44.62 & 50.86 & 56.91 & 62.81 & 68.56  \\
$H^{10}_n$     & 10    & 19.00 & 27.43 & 35.44 & 43.13 & 50.57 & 57.78 & 64.81 & 71.67 & 78.38  \\
$H^{15}_n$     & 15    & 29.00 & 42.42 & 55.42 & 68.09 & 80.51 & 92.70 & 104.7 & 116.5 & 128.2  \\
$H^{20}_n$     & 20    & 39.00 & 57.46 & 75.42 & 93.09 & 110.5 & 127.7 & 144.7 & 161.5 & 178.2  \\
$\lg(n!)$  & 0     & 1     & 2.585 & 4.585 & 6.907 & 9.492 & 12.30 & 15.30 & 18.47 & 21.79  \\
$D_n$        & 0     & 2     & 2.667 & 3.143 & 3.505 & 3.794 & 4.035 & 4.241 & 4.421 & 4.581  \\
$F_n$        & 1     & 0.667 & 0.714 & 0.724 & 0.724 & 0.722 & 0.721 & 0.721 & 0.721 & 0.721 \\
\hline\hline
$n$&20&50&100&200&500&1000&2000&5000&10000&100000\\\hline\hline
$h_n$        & 3.207 & 3.869 & 4.369 & 4.869 & 5.530 & 6.030 & 6.530 & 7.191 & 7.691 & 9.352 \\
$H'_n$       & 42.45 & 111.8 & 227.9 & 460.7   & 1160   & 2327 & 4658 & 11656 & 23319 & 233264 \\
$H_n$        & 51.29 & 133.9 &  272.2 &  549.2 &  1381  & 2768 & 5543 & 13869 & 27746 & 277534 \\
$H^5_n$      & 58.82 & 136.1 &  272.3 &  549.2 &  1381  & 2768 & 5543 & 13869 & 27746 & 277534 \\
$H^9_n$      & 120.4 & 245.1 &  411.6 &  692.1 &  1462  & 2789 & 5544 & 13869 & 27746 & 277534 \\
$H^{10}_n$     & 139.7 & 290.5 & 494.0 &  827.6 & 1651  & 2931 & 5584 & 13869 & 27746 & 277534 \\
$H^{15}_n$     & 238.9 &  535.9 & 975.8 & 1757 &  3748 &  6531 & 11186 & 22219 & 37075 & 277758 \\
$B_n(F^{15}_n)$ & 308.1 &  675.0 & 1206  & 2124 & 4363  & 7305  & 11809 & 20608 & 28813 & 69971 \\
$H^{20}_n$     & 338.9 & 785.8 & 1475 & 2755  &  6233  & 11473 & 20956 & 45815 & 81732 & 501779 \\
$B_n(F^{20}_n)$ & 452.4 & 1036  & 1927 & 3565  & 7960  & 14478  & 26073 & 55666 & 96970 & 502238 \\
$H^{30}_n$     & 538.9 & 1286  & 2475 & 4755  & 11233  & 21471 & 40947 & 95768 & 181542 & 1483314 \\
$B_n(F^{30}_n)$ & 740.9 & 1757  & 3370 & 6451  & 15173  & 28903 & 54921 & 127767& 241107 & 1931833 \\
$\lg(n!)$  & 61.08 & 214.2 & 524.8 &  1245  &  3767  &  8529 & 19053 & 54233 & 118458 & 1516704 \\
$D_n$        & 5.62 & 6.962 & 7.969  &  8.973 &  10.30 &  11.30 & 12.30 & 13.62 & 14.62 & 17.94\\
$F^9_n$      & 0.038 & 0.092 & 0.172 & 0.304  & 0.542  &  0.681 & 0.720 & 0.721 & 0.721 & 0.721 \\
$F^{10}_n$   & 0.019& 0.047 & 0.092  & 0.172  & 0.358  &  0.542 & 0.680 & 0.721 & 0.721 & 0.721 \\
$10^2 F^{15}_n$ &0.061& 0.152& 0.305 & 0.608  & 1.510  &  2.991 & 5.862 & 13.80 & 25.05 & 71.45 \\
$10^5 F^{20}_n$ &1.907&4.768& 9.536  & 19.07  & 47.67  &  95.31 & 190.5 & 475.3 & 947.6 & 8954 \\
$10^8 F^{30}_n$ &1.863&4.657& 9.313  & 18.63  & 46.57  &  93.13 & 186.3 & 465.7 & 931.3 & 9313\\
\hline
\end{tabular}
\end{table*}
\section*{Acknowledgment}
I would like to thank Witek Baryluk for bringing Bloom filters to my attention and discussion and James Dow Allen for pointing out reduction of the minimal prefix tree and discussion.

\bibliographystyle{plain}

\end{document}